\documentclass[english,11pt,a4paper]{article}
\usepackage{jheppub}
\usepackage[T1]{fontenc}
\usepackage[utf8]{inputenc}
\usepackage{float}
\usepackage{graphicx, hyperref}
\usepackage{amssymb,amsmath}
\usepackage{esint}
\usepackage{epstopdf}
\usepackage{babel}

\newcommand{\eq}{\begin{equation}}
\newcommand{\eqx}{\end{equation}}
\newcommand{\eqs}{\begin{equation*}}
\newcommand{\eqsx}{\end{equation*}}
\newcommand{\eqn}{\begin{eqnarray}}
\newcommand{\eqnx}{\end{eqnarray}}
\newcommand{\alg}{\begin{align}}
\newcommand{\algx}{\end{align}}

\newcommand{\f}[2]{\frac{#1}{#2}}

\newcommand{\sg}{\sigma}

\newcommand{\Dl}{\Delta}

\newcommand{\qqqq}{\quad\quad\quad\quad}

\newcommand{\nn}{{\cal N}}

\newcommand{\eloop}{E_{direct}}
\newcommand{\emod}{E_{mod}}

\begin{document}

\title{  Six and seven loop Konishi from L{\"u}scher corrections}

\author[a]{Zolt\'an Bajnok}
\author[b]{and Romuald A. Janik}

\affiliation[a]{MTA Lend\"ulet Holographic QFT Group,
Wigner Research Centre,\\
H-1525 Budapest 114, P.O.B. 49, Hungary}
\affiliation[b]{Institute of Physics,
Jagiellonian University,\\
ul. Reymonta 4, 30-059 Krak\'ow, Poland} 

\emailAdd{bajnok.zoltan@wigner.mta.hu}
\emailAdd{romuald@th.if.uj.edu.pl}

\abstract{
In the present paper we derive six and seven loop formulas
for the anomalous dimension of the Konishi operator in N=4 SYM
from string theory using the technique of L\"uscher corrections.
We derive analytically the integrand using the worldsheet S-matrix
and evaluate the resulting integral and infinite sum
using a combination of high precision numerical integration
and asymptotic expansion. We use this high precision
numerical result to fit the integer coefficients
of zeta values in the final analytical answer. 
The presented six and seven loop results can be
used as a cross-check with FiNLIE on the string theory side,
or with direct gauge theory
computations. The seven loop level is the
theoretical limit of this L\"uscher approach
as at eight loops double-wrapping corrections will
appear.}

\maketitle

\newpage

\section{Introduction}

The Konishi operator is the simplest nonprotected local operator in
$\nn=4$ Super-Yang-Mills theory. For this reason, its properties serve
as a theoretical laboratory in our quest for the complete solution
of $\nn=4$ SYM at any coupling. The relative simplicity of its structure
and properties allows us to probe them using various methods -- including
direct gauge theoretical computations and computations on the string side
of the AdS/CFT correspondence. On both sides of the AdS/CFT duality there
are various ways (sometimes conceptually quite different) of performing
these computations and the agreement of the computations was always
a strong confirmation of the respective methods.
Apart from the above mentioned comparisons between different computations
for the Konishi operator, there are also sometimes some quite nontrivial
internal consistency checks within one particular method (e.g. the cancellation
of dynamical pole contributions in the L{\"u}scher correction method,
the general transcendentality structure of the final result) which 
serve as a very strong and nontrivial constraint on the ingredients
entering the computation like the fine structure of the S-matrix and
its most complicated component -- the so-called `dressing phase' etc.

The anomalous dimension of the Konishi operator up to three loops is
beautifully described by the Asymptotic Bethe Ansatz (ABA) of Beisert and Staudacher
within the spin-chain approach \cite{BS}.

The four-loop anomalous dimension of the Konishi operator was particularly
important as it was only starting from this order of perturbation theory
that deviations from the ABA appear due to a new topological class of
Feynman diagrams -- so-called `wrapping interactions'. These effects
cannot be understood within the spin-chain point of view and their
description neccessitates the use of the string $\sg$-model in 
$AdS_5 \times S^5$. The `wrapping corrections' are then naturally
identified \cite{AJK} with finite-size L{\"u}scher virtual corrections
charactersitic of a two-dimensional worldsheet QFT. The agreement
of direct gauge theory perturbative computations \cite{FSSZ,Velizhanin08} and the
string $\sg$-model worldsheet QFT computation \cite{Bajnok:2008bm} showed
effectively that the gauge theory spin chain breaks down 
exactly in the way expected from a two-dimensional QFT on a cylinder
i.e. string theory.

The five-loop order was computed initially only from
the string L{\"u}scher corrections~\cite{Bajnok:2009vm}. The main motivation for this
computation was that at five loops many more nontrival ingredients
of the string $\sg$-model computation start to appear, in particular
an infinite set of coefficients of the dressing phase and virtual
modifications of the quantization conditions for the constituents
of the Konishi state. Here the test of these ingredients were initially
just the internal cross-checks of the L{\"u}scher computation like the
cancellation of so-called dynamical pole contributions and a simple
transcendental structure of the final result. A second motivation for
this computation was to provide a cross-check for the exact sets of
Thermodynamic Bethe Ansatz equations proposed for the spectrum
of $\nn=4$ SYM theory \cite{TBA1,TBA2,TBA3}. Subsequently full agreement was found,
first numerically \cite{Arutyunov:2010gb} and 
then analytically~\cite{Balog:2010xa}.

It is worth mentioning that the Konishi operator is the first member of an
important family of so-called twist-two operators, whose anomalous dimensions
obey the still mysterious maximal transcendentality principle \cite{transcend}
and intricate relations with BFKL. In \cite{KLRSV} it was found that
the Asymptotic Bethe Ansatz answer is inconsistent with BFKL starting
from four loop level, with the discrepancy being attributed to the
then unknown wrapping corrections.
The wrapping corrections for these
operators have been computed on the string side using L{\"u}scher corrections
at four- and five-loop level \cite{fourlooptwisttwo,fivelooptwisttwo}, 
restoring consistency with BFKL expectations.

L{\"u}scher corrections for multiparticle states include only the single
wrapping contribution. However
the sets of TBA equations \cite{TBA1,TBA2,TBA3} proposed later, provide 
a method of obtaining the exact
anomalous dimension for any coupling thus resumming all multiple 
wrapping effects at once. Numerical computations of the Konishi anomalous
dimension for a wide range of couplings have been performed in \cite{Gromov:2009zb} and 
confirmed in \cite{Frolov:2010wt}. Subsequently large coupling asymptotic fits 
to the numerical
data have been shown to agree with string theory computations 
\cite{Gromov:2011de,Roiban:2011fe,Vallilo:2011fj}.

The reason for returning to the six- and seven-loop Konishi anomalous dimension 
using the L{\"u}scher correction approach in the present
paper has been twofold. Firstly, on the gauge theory side new techniques
have been developed which allowed for a direct gauge theory computation
at the five-loop level \cite{Eden:2012fe} and the verification of the previous 
string computation \cite{Bajnok:2009vm}. The authors informed us that a 
six-loop gauge theory computation
using their methods was within reach and expressed interest in having
a string theory cross-check. However, numerical methods of solving TBA
equations are not precise enough to extract such high loop level coefficients,
thus making the application of the conventional L{\"u}scher computation
an attractive alternative when applicable.

Secondly, the infinite set of TBA equations has been recast in a form
of a finite set of nonlinear integral equations (FiNLIE) \cite{NLIE1,NLIE2}, 
which makes a weak coupling
expansion practically possible -- however it is still very complex.
This endeavor is now almost finalized \cite{LSV}, and the present 
L{\"u}scher computation
may provide a nontrivial cross-check.

For the above reasons we have embarked on computing the six and seven loop
anomalous dimension of the Konishi operator using the multiparticle 
L{\"u}scher correction formulas of \cite{Bajnok:2008bm,Bajnok:2009vm}. 
The seven loop level is the
theoretical limit of this approach as at eight loops double-wrapping will
appear which is not taken into account by these formulas.

The general structure of the result of the multiparticle 
L{\"u}scher correction formula is
\eq
\Dl E_{6,7-loop} = -\sum_{Q=1}^\infty \f{1}{2\pi} \int_{-\infty}^\infty dq \, 
f_{6,7-loop}(q,Q) 
\eqx  
where the integrand can be obtained in a relatively straightforward way.
At $n$-loop the integrand contains the $(n-4)^{th}$ power  of the polygamma function
originating from the dressing phase.
Evaluating the integral by residues  \cite{Bajnok:2009vm,Bajnok:2010ud}
derivatives of the polygamma function at integers values appear which, after 
summation for $Q$, can be expressed in terms of multiple zeta values (MZVs). 
At $n$-loop MZVs of depth $n-3$ appear \cite{Bajnok:2010ud}. Interestingly, 
irreducible MZVs showing up at various stages of the calculations cancel completely
leaving only products of simple zeta functions. We do not have an understanding of this 
phenomenon yet, but will assume that similar structure will appear for 
the Konishi operator, too. Of course, the fact that our final results 
can be fitted to a very high accuracy without using irreducible MZV's
strongly supports this assumption.

In the present case we have decided to bypass the technical steps above, 
as at six and seven loops
they would be prohibitively complicated. We derive just the analytical
expression for the integrand $f(q,Q)$ and perform the resulting integral
and sum semi-numerically to a very high precision\footnote{This is akin
to the method used in \cite{fivelooptwisttwo}. In our case, however,
a purely numerical summation does not suffice to attain the required
accuracy and we use several orders of additional asymptotic analysis.}
 which allows us to
determine the coefficients of the zeta-functions in the expected
form of the result
\eq
\Dl E_{6-loop} = c_1+c_2\, \zeta(3)+c_3\, \zeta(5)+c_4\, \zeta(7)+c_5\, \zeta(9)+
c_6\, \zeta^2(3)+c_7\, \zeta(3)\zeta(5)
\eqx
and
\eqn
\Dl E_{7-loop} &=& c_1+c_2\, \zeta(3)+c_3\, \zeta(5)+c_4\, \zeta(7)+c_5\, \zeta(9)+
c_6\, \zeta(11)+ c_7\, \zeta^2(3)+ \nonumber\\
&&c_8\, \zeta(3)\zeta(5) +c_9\, \zeta(3)\zeta(7)+
c_{10}\, \zeta^3(3) +c_{11}\, \zeta^2(5)
\eqnx
assuming, as was the case for lower orders, that the coefficients are integers.

The plan of the paper is as follows. The anomalous dimension of 
the Konishi operator corresponds to the finite volume energy of a two 
particle scattering state.  In section 2 we derive the ABA
momentum and energy of this state up to seven loops. In Section 3 we express
the leading vacuum polarization effects in terms of the scattering matrix and 
evaluate their weak coupling expansion, thus obtaining the
integrand of the multiparticle L{\"u}scher correction formula. 
These expressions are subject to a 
consistency check in section~4, where we also compute analytically the highest
transcendental part of the answer which will be an important cross-check
for our numerical approach. In section 5 we discuss the semi-numerical 
computation of our result. Finally we provide the full dimension of 
the Konishi operator up to seven loops in section 6 and make some 
concluding remarks. 
 
\section{Asymptotic Bethe Ansatz up to 7-loops}

Finite volume energy corrections of multiparticle states have two
sources: momentum quantization and vacuum polarization. Asymptotic
BA determines momentum quantization and captures all finite size corrections,
which are polynomial in the inverse of the volume. In contrast,  vacuum polarization
effects are exponentially supressed and their leading term is described
by L\"uscher-type formulas. 

The Konishi operator corresponds to a two-particle state in volume
$L=2$. These particles scatter diagonally on each other and the asymptotical
BA equation quantizes their momenta as:
\begin{equation}
e^{ipL}=S(p,-p)\,,\label{eq:BY}
\end{equation}
Once the momenta are determined the asymptotic energy of the state
can be calculated from the dispersion relation%
\footnote{The 't Hooft coupling is reletad to $g$ as $g=\frac{\sqrt{\lambda}}{4\pi}$.%
}
\begin{equation}
\epsilon(p)=\sqrt{1+16g^{2}\sin^{2}\frac{p}{2}}
\end{equation}
as 
\[
E_{ABA}(p)=\epsilon(p)+\epsilon(-p)=2\epsilon(p).
\]
The scattering matrix takes the form: 
\begin{equation}
S(p,-p)=\frac{\epsilon(p)\cot\frac{p}{2}+i}{\epsilon(p)\cot
\frac{p}{2}-i}e^{-2ip}e^{2i\theta(p,-p)}\,,
\end{equation}
where the dressing phase, $\theta(p,-p)$, is parameterized in the
following manner:
\begin{equation}
\theta(x_{1},x_{2})=\chi(x_{1}^{+},x_{2}^{+})+\chi(x_{1}^{-},x_{2}^{-})-
\chi(x_{1}^{+},x_{2}^{-})-\chi(x_{1}^{-},x_{2}^{+})
\end{equation}
with
\begin{equation}
x^{\pm}(p)=\frac{1}{4g}(\cot\frac{p}{2}\pm i)(1+\epsilon(p))
\end{equation}
The functions $\chi$ have a double expansion 
\begin{equation}
\chi(x_{1},x_{2})=-\sum_{r=2}^{\infty}\sum_{s>r}\frac{c_{r,s}(g)}{(r-1)(s-1)}
\left[\frac{1}{x_{1}^{r-1}x_{2}^{s-1}}-\frac{1}{x_{1}^{s-1}x_{2}^{r-1}}\right]
\end{equation}
and the $c_{r,s}$ coefficients, 
\begin{equation}
c_{r,s}(g)=(r-1)(s-1)2\cos(\frac{\pi}{2}(s-r-1))\int_{0}^{\infty}dt
\frac{J_{r-1}(2gt)J_{s-1}(2gt)}{t(e^{t}-1)},
\end{equation}
 are written such that in the weak coupling limit ($g\to0$) the integrals can be easily
evaluated to yield: 
\begin{eqnarray}
c_{2,3}(g) & = & 4\zeta_{3}g^{3}-40\zeta_{5}g^{5}+420\zeta_{7}g^{7}-4704\zeta_{9}g^{9}+\dots\\
c_{2,5}(g) & = & -8\zeta_{5}g^{5}+168\zeta_{7}g^{7}+\dots\nonumber \\
c_{3,4}(g) & = & 24\zeta_{5}g^{5}-420\zeta_{7}g^{7}+\dots\nonumber 
\end{eqnarray}
We solve the asymptotic BA equation (\ref{eq:BY}) iteratively: 
\begin{eqnarray}
p =  &+&\frac{2\pi}{3}-\sqrt{3}g^{2}+\frac{9}{2}\sqrt{3}g^{4}-24\sqrt{3}(1+\zeta_{3})g^{6}
+\frac{\sqrt{3}}{4}(671+960(\zeta_{3}+\zeta_{5}))g^{8}\\
 & - & \frac{6\sqrt{3}}{5}(1269+1880\zeta_{3}+2060\zeta_{5}+2100\zeta_{7})g^{10}\nonumber \\
 & + & 6\sqrt{3}(2643+12\zeta_{3}(311+8\zeta_{5})+4072\zeta_{5}+4452\zeta_{7}+4704
    \zeta_{9})g^{12} \nonumber 
\end{eqnarray}
By plugging back this asymptotic momentum into the dispersion relation
we obtain the asymptotic energy of the Konishi state: 
\begin{eqnarray}
\label{e.ABA}
E_{BY}=2\epsilon(p) & = & 2+12g^{2}-48g^{4}+336g^{6}-12(235+24\zeta_{3})g^{8}\\
 &  & +12(2209+360\zeta_{3}+240\zeta_{5})g^{10}\nonumber \\
 &  & -12(22429+4608\zeta_{3}+3672\zeta_{5}+2520\zeta_{7})g^{12}\nonumber \\
 &  & +24(119885+24\zeta_{3}(1211+6\zeta_{3})+24156\zeta_{5}+19656\zeta_{7}+14112\zeta_{9})g^{14}\nonumber 
\end{eqnarray}
This provides the exact answer up to $g^{6}$ (three loops) as from
four loops onwards wrapping corrections start to contribute.

\section{L\"uscher type corrections}

The vacuum polarization effects show up in two different ways. First,
particles polarize the vacuum in which they are moving, this is what
we call the direct energy correction. Second, the polarized vacuum
reacts back and modifies -- via the asymptotic Bethe Ansatz -- the particles
momenta, this we call the BA modification. Let us analyze them one after
one the other.

\subsection{Direct energy correction}

The polarized vacuum will contribute to the energy as 
\begin{equation}
\Delta E=-\sum_{Q=1}^{\infty}\int\frac{dq}{2\pi}
\mbox{sTr}(S_{Q1}^{Q1}(q,p)S_{Q1}^{Q1}(q,-p))e^{-\tilde{\epsilon}_{Q}(q)L}
+O(g^{16})\label{eq:Luscher}
\end{equation}
Here we sum up for all bound-states of charge $Q$ of the mirror model,
whose momenta are $q$. $S_{Q1}^{Q1}$ describes how these mirror
bound-states scatter on the fundamental ``Konishi'' particle, while
$\tilde{\epsilon}_{Q}(q)$ denotes their mirror energy. This expression
is exact in the weak coupling limit up to the order $g^{14}$ since
at the order $g^{16}$ double wrapping effects will contribute, too.
In order to obtain the six- and seven-loop anomalous dimension of the Konishi
operator we need to expand this expression up to $g^{14}.$ Let us
analyze the $g$ dependence of the various terms.

\subsubsection*{Exponential factor}

The mirror energy has the following parametrization
\begin{equation}
e^{-\tilde{\epsilon}_{Q}(q)}=\frac{z^{-}(q,Q)}{z^{+}(q,Q)}\quad;
\qquad z^{\pm}(q,Q)=\frac{q+iQ}{4g}\left(\sqrt{1+\frac{16g^{2}}{q^{2}+Q^{2}}}\pm1\right)
\end{equation}

\subsubsection*{Matrix part of the scattering matrix}

There are many representative of the Konishi operator and in the present
paper we choose the $su(2)$ one which we label by $\left(1\dot{1}\right)$.
This choice has the advantage that only diagonal matrix elements will
contribute to eq. (\ref{eq:Luscher}). The scattering matrix can be factorized as 
\begin{equation}
S_{(\alpha\dot{\alpha})(1\dot{1})}^{(\alpha\dot{\alpha})(1\dot{1})}(z^{\pm},x^{\pm})
=S_{scal}(z^{\pm},x^{\pm})S_{\alpha1}^{\alpha1}(z^{\pm},x^{\pm})
S_{\dot{\alpha}\dot{1}}^{\dot{\alpha}\dot{1}}(z^{\pm},x^{\pm})
\end{equation}
where $(\alpha,\dot{\alpha})$ labels the polarization of the mirror
bound-states. For one single $su(2\vert2)$ algebra the basis for
boundstates can be represented in the superspace formalism as $(w_{3}^{j}w_{4}^{Q-j}$,
$w_{3}^{j}w_{4}^{Q-2-j}\theta_{1}\theta_{2}$, $w_{3}^{j}w_{4}^{Q-1-j}\theta_{1}$,
$w_{3}^{j}w_{4}^{Q-1-j}\theta_{2})$, where $j$ takes values from
$0$ to $(Q,Q-2,Q-1,Q-1)$, respectively. See \cite{Arutyunov:2008zt,Bajnok:2008bm}
for the details. As the two $su(2\vert2)$ factors are equivalent
the contribution of the scattering matrix can be factored as 
\begin{equation}
\mbox{sTr}(S_{Q1}^{Q1}(q,p)S_{Q1}^{Q1}(q,-p)) =   S_{scal}(q,p)S_{scal}(q,-p)
 \mbox{sTr}(S_{mat}^{su(2)}(q,p)S_{mat}^{su(2)}(q,-p))^{2} 
\end{equation}
 We use formulas from \cite{Bajnok:2008bm} and the following identities:
\begin{equation}
S_{\alpha1}^{\alpha1}(z^{\pm},x^{\pm})=S_{1\alpha}^{1\alpha}(x^{\mp},z^{\mp})
=S_{3\bar{\alpha}}^{3\bar{\alpha}}(x^{\pm},z^{\pm})
\end{equation}
where $\bar{\alpha}$ is the image of $\alpha$ under the transformation
$1\leftrightarrow3$, $2\leftrightarrow4$, but it will not be important
here as in the supertrace we sum over all states. From \cite{Bajnok:2008bm}
we can extract the matrix elements, which turn out to be independent
of $j$:
\begin{equation}
w_{1}^{j}w_{2}^{Q-j}\to a_{5}^{5}(x,z) \equiv S1(z,x^{\pm})
=\frac{z^{+}-x^{+}}{z^{-}-x^{+}}\frac{\tilde{\eta}_{1}}{\eta_{1}}
\end{equation}
\begin{equation}
w_{1}^{j}w_{2}^{Q-2-j}\theta_{3}\theta_{4}\to 
2a_{8}^{8}(x,z) \equiv S2(z,x^{\pm})=\frac{z^{+}-x^{-}}{z^{-}-x^{+}}
\frac{(1-x^{+}z^{-})}{(1-x^{-}z^{-})}\frac{x^{-}}{x^{+}}\frac{\tilde{\eta}_{1}}{
\eta_{1}}\left(\frac{\tilde{\eta}_{2}}{\eta_{2}}\right)^{2}
\end{equation}
\begin{equation}
w_{1}^{j}w_{2}^{Q-1-j}\theta_{3}\to a_{9}^{9}(x,z)
\equiv S3(z,x^{\pm})=\frac{z^{+}-x^{-}}{z^{-}-x^{+}}
\frac{\tilde{\eta}_{1}}{\eta_{1}}\frac{\tilde{\eta}_{2}}{\eta_{2}}
\end{equation}
\begin{equation}
w_{1}^{j}w_{2}^{Q-1-j}\theta_{4}\to 
\frac{1}{2}(a_{3}^{3}(x,z)+a_{9}^{9}(x,z)) \equiv S4(z,x^{\pm})=
\frac{z^{+}-x^{+}}{z^{-}-x^{+}}\frac{(1-x^{+}z^{-})}{(1-x^{-}z^{-})}
\frac{x^{-}}{x^{+}}\frac{\tilde{\eta}_{1}}{\eta_{1}}\frac{\tilde{\eta}_{2}}{\eta_{2}}
\end{equation}
The string frame factors appearing here can be written as $\frac{\tilde{\eta}_{1}}{\eta_{1}}
=\sqrt{\frac{z^{-}}{z^{+}}}$
and $\left(\frac{\tilde{\eta}_{2}}{\eta_{2}}\right)^{2}=\frac{x^{+}}{x^{-}}$.
As in the Konishi state we have two particles with opposite momenta
the factors $\frac{\tilde{\eta}_{2}}{\eta_{2}}$ will cancel:
\begin{eqnarray}
\mbox{sTr}(S_{mat}^{su(2)}(z,x^{\pm})S_{mat}^{su(2)}(z,-x^{\mp})) =  
 (Q+1)S1(z,x^{\pm})S1(z,-x^{\mp}) \phantom{aaaaaaaaaaaaaaaaaaaa} & & \\ 
 +(Q-1)S2(z,x^{\pm})
S2(z,-x^{\mp}) -Q S3(z,x^{\pm})S3(z,-x^{\mp})- Q
S4(z,x^{\pm})S4(z,-x^{\mp})& & \nonumber \\
  =  \frac{2(x^{-}-x^{+})(z^{+}(-1-2Qx^{+}z^{-}+z^{-}z^{+})+x^{-}z^{-}(2Qz^{+}
+x^{+}(-1+z^{-}z^{+})))z^{-}}{(x^{-}+z^{-})(z^{-}-x^{+})(-1+x^{-}z^{-})(1+x^{+}z^{-})z^{+}} 
\phantom{aaa} & &\nonumber
\end{eqnarray}

\subsubsection*{Scalar factor}

The scalar factor for the fundametal particles is 
\begin{equation}
S_{scal}^{Q=1}(z,x)=\frac{z^{-}-x^{+}}{z^{+}-x^{-}}
\frac{1-\frac{1}{z^{+}x^{-}}}{1-\frac{1}{z^{-}x^{+}}}
\end{equation}
and this is what we should fuse for the bound-state:
\begin{equation}
S_{scal}(z,x)=\prod_{i}S_{scal}^{Q=1}(z_{i},x)
\end{equation}
There is a nicer form, however, which does not depend on the individual
constituents:
\begin{equation}
S_{scal}^{-1}(z,x)=\Sigma_{Q,1}^{2}(z,x)S^{su(2)}(z,x)
\end{equation}
The $su(2)$ scalar factor is defined as: 
\begin{equation}
S^{su(2)}(z,x)=\frac{(z^{+}-x^{-})(z^{+}-x^{+})}{(z^{-}-x^{+})(z^{-}-x^{-})}
\frac{(1-\frac{1}{z^{+}x^{-}})(1-\frac{1}{z^{+}x^{+}})}{(1-\frac{1}{z^{-}x^{+}})(1-
\frac{1}{z^{-}x^{-}})}
\end{equation}
while for the case $\vert x^{\pm}\vert>1$ following \cite{Arutyunov:2009kf}
we can write%
\footnote{We have to be careful as the conventions of \cite{Arutyunov:2009kf}
are different from ours. To turn into our conventions we made the
replacements: $z^{\pm}\to z^{\mp}$ and $x^{\pm}\to x^{\mp}.$%
} 
\begin{eqnarray}
-i\log\Sigma_{Q,1}(z,x) & = & \Phi(z^{+},x^{+})-\Phi(z^{+},x^{-})-\Phi(z^{-},x^{+})
+\Phi(z^{-},x^{-})\\
 &  & -\frac{1}{2}\left[-\Psi(z^{+},x^{+})+\Psi(z^{+},x^{-})-\Psi(z^{-},x^{+})
+\Psi(z^{-},x^{-})\right]\nonumber \\
 &  & +\frac{1}{2i}\log\left[\frac{(z^{-}-x^{-})(x^{+}-\frac{1}{z^{-}})^{2}}{(z^{-}
-x^{+})(x^{+}-\frac{1}{z^{+}})(x^{-}-\frac{1}{z^{+}})}\right]\nonumber 
\end{eqnarray}
The functions $\Phi$ and$\Psi$ are defined in terms of the integrals 
\begin{equation}
\Phi(x_{1},x_{2})=i\oint_{C_{1}}\frac{dw_{1}}{2\pi i}\oint_{C_{1}}\frac{dw_{2}}{2\pi i}
\frac{1}{w_{1}-x_{1}}\frac{1}{w_{2}-x_{2}}\log\frac{\Gamma(1+ig(w_{1}
+w_{1}^{-1}-w_{2}-w_{2}^{-1}))}{\Gamma(1-ig(w_{1}+w_{1}^{-1}-w_{2}-w_{2}^{-1}))}
\end{equation}
\begin{equation}
\Psi(x_{1},x_{2})=i\oint_{C_{1}}\frac{dw_{2}}{2\pi i}\frac{1}{w_{2}-x_{2}}\log\frac{
\Gamma(1+ig(x_{1}+x_{1}^{-1}-w_{2}-w_{2}^{-1}))}{\Gamma(1-ig(x_{1}+x_{1}^{-1}-w_{2}-w_{2}^{-1}))}
\end{equation}
We separate the rational part from the $\Psi$ and $\Phi$ functions
as 
\begin{equation}
S_{scal}(z,x)=S_{0}(z,x)e^{i\sigma(z,x)}
\end{equation}
where 
\begin{equation}
S_{0}(z,x)=\frac{(z^{-}-x^{+})^{2}(1-z^{-}x^{-})}{(z^{+}-x^{-})(z^{+}-x^{+})(1-z^{-}x^{+})}
\end{equation}
and 
\begin{eqnarray}
\sigma(z,x) & = & 2(\Phi(z^{+},x^{+})-\Phi(z^{+},x^{-})-\Phi(z^{-},x^{+})+\Phi(z^{-},x^{-}))\\
 &  & -(-\Psi(z^{+},x^{+})+\Psi(z^{+},x^{-})-\Psi(z^{-},x^{+})+\Psi(z^{-},x^{-}))\nonumber 
\end{eqnarray}

\subsubsection*{Weak coupling expansion}

Our aim is to calculate the weak coupling expansion of $\Delta E$
for $L=2$. In so doing we decompose the integrand of the L\"uscher
correction (\ref{eq:Luscher}) 
\begin{equation}
\Delta E=-\sum_{Q=1}^{\infty}\int\frac{dq}{2\pi}P(q,Q)\Sigma(q,Q)
\end{equation}
into a simpler rational part 
\begin{equation}
P(q,Q)=\frac{4(x^{-}-x^{+})^{2}(z^{+}(-1-2Qx^{+}z^{-}+z^{-}z^{+})+x^{-}z^{-}(2Qz^{+}+x^{+}
(-1+z^{-}z^{+})))^{2}}{((z^{+})^{2}-(x^{+})^{2})((z^{+})^{2}-(x^{-})^{2})(-1+(x^{-})^{2}
(z^{-})^{2})(1+(x^{+})^{2}(z^{-})^{2})}\left(\frac{z^{-}}{z^{+}}\right)^{4}
\end{equation}
which contains both the matrix part and the rational part of the scalar
factor, and into the more complicated $\Sigma$ part: 
\begin{eqnarray}
i\log\Sigma(q,Q) & = & -2(\Phi(z^{+},x^{+})-\Phi(z^{+},x^{-})-\Phi(z^{-},x^{+})+
\Phi(z^{-},x^{-}))\\
 &  & -2(\Phi(z^{+},-x^{-})-\Phi(z^{+},-x^{+})-\Phi(z^{-},-x^{-})+\Phi(z^{-},-x^{+}))\nonumber \\
 &  & -\Psi(z^{+},x^{+})+\Psi(z^{+},x^{-})-\Psi(z^{-},x^{+})+\Psi(z^{-},x^{-})\nonumber \\
 &  & -\Psi(z^{+},-x^{-})+\Psi(z^{+},-x^{+})-\Psi(z^{-},-x^{-})+\Psi(z^{-},-x^{+})\nonumber 
\end{eqnarray}
We expand these functions in $g^{2}$ as 
\begin{equation}
P(q,Q)=P_{8}(q,Q)g^{8}+P_{10}(q,Q)g^{10}+P_{12}(q,Q)g^{12}+P_{14}(q,Q)g^{14}+\dots
\end{equation}
\begin{equation}
\Sigma(q,Q)=1+\Sigma_{2}(q,Q)g^{2}+\Sigma_{4}(q,Q)g^{4}+\Sigma_{6}(q,Q)g^{6}+\dots
\end{equation}
The expansion of the rational part is quite straightforward. In expanding
the $\Psi$ and $\Phi$ functions we use the same method we used in
\cite{Bajnok:2009vm}. The expansion of the $\Psi(x_{1},x_{2})$ functions
for $\vert x_{2}\vert>1$ (string region) reads as follows 
\begin{eqnarray}
\Psi(x_{1},x_{2}) & = & -\frac{g}{x_{2}}(\Psi(1-ig(x_{1}+x_{1}^{-1}))+\Psi(1+ig(x_{1}+x_{1}^{-1})))\\
 &  & -\frac{ig^{2}}{2x_{2}^{2}}(\Psi_{1}(1-ig(x_{1}+x_{1}^{-1}))-\Psi_{1}(1+ig(x_{1}+x_{1}^{-1})))\nonumber \\
 &  & +g^{3}\Bigl(\frac{1}{2x_{2}}+\frac{1}{6x_{2}^{3}}\Bigr)(\Psi_{2}(1-ig(x_{1}+x_{1}^{-1}))+\Psi_{2}(1+ig(x_{1}+x_{1}^{-1})))\nonumber \\
 &  & +\frac{ig^{4}}{6x_{2}^{2}}(\Psi_{3}(1-ig(x_{1}+x_{1}^{-1}))-\Psi_{3}(1+ig(x_{1}+x_{1}^{-1})))\nonumber \\
 &  & -\frac{g^{5}}{12x_{2}}(\Psi_{4}(1-ig(x_{1}+x_{1}^{-1}))+\Psi_{4}(1+ig(x_{1}+x_{1}^{-1})))+\dots\nonumber 
\end{eqnarray}
where $\Psi_{n}(x)=(\frac{d}{dx})^{n+1}(\log(\Gamma(x))$ are the
standard polygamma functions. If $\vert x_{1}\vert>1$ then $\Phi(x_{1},x_{2})$
starts at $g^{6}$ and coincides with $\chi(x_1,x_2)$. In the opposite case using the identity 
$\Phi(x_{1},x_{2})=\Phi(0,x_{2})-\Phi(x_{1}^{-1},x_{2})$,
being valid if $\vert x_{1}\vert\neq1$, we can calculate the leading
expansion of $\Phi$ as 
\begin{equation}
\Phi(0,x)=\frac{2\gamma_{E}g}{x}-2\zeta_{3}g^{3}\left(\frac{3}{x}+\frac{1}{3x^{3}}\right)+24\zeta_{5}g^{5}\left(\frac{5}{3x}+\frac{1}{3x^{3}}\right)+\dots
\end{equation}
\begin{equation}
\Phi(x_{1}^{-1},x_{2})=-2\zeta_{3}g^{3}\left(\frac{x_{1}}{x_{2}^{2}}-\frac{x_{1}^{2}}{x_{2}}\right)+\dots
\end{equation}
 The $\Psi$ terms simplify, too: 
\begin{eqnarray}
& & i\log\Sigma(q,Q) = \\  &  &2g(\frac{1}{x^{+}}-\frac{1}{x^{-}})\biggl(4\gamma_{E} 
+\Psi(1-igu^{+}) +\Psi(1+igu^{+})+\Psi(1-igu^{-})+\Psi(1+igu^{-})\biggr) \nonumber \\
 &  & -g^{3}(\frac{1}{x^{+}}-\frac{1}{x^{-}})\biggl(8\,\zeta_{3}\cdot \left(3+\frac{1}{(z^{+})^{2}}+(z^{-})^{2}\right)
\nonumber \\
& & \phantom{aaaaaaaaaaaaaaa}+\Psi_{2}(1-igu^{+})+\Psi_{2}(1+igu^{+})+\Psi_{2}(1-igu^{-})+\Psi_{2}(1+igu^{-})\biggr)\nonumber \\
 &  & -\frac{g^{5}}{6}(\frac{1}{x^{+}}-\frac{1}{x^{-}})
\biggl(960\,\zeta_{5} +\Psi_{4}(1-igu^{+})
+\Psi_{4}(1+igu^{+})+\Psi_{4}(1-igu^{-})+\Psi_{4}(1+igu^{-})\biggr) \nonumber
\end{eqnarray}
where 
\begin{equation}
u^{\pm}=z^{\pm}+\frac{1}{z^{\pm}}
\end{equation}
We have to plug back the parametrization of $x^{\pm}$ and $z^{\pm}$
and expand in $g$ to obtain $\Sigma_{2}$, $\Sigma_{4}$ and $\Sigma_{6}$. This
leads to an explicit, however very complicated expression for the
direct energy integrand in the form 
\begin{equation}
P_{12}(q,Q)+P_{10}(q,Q)\Sigma_{2}(q,Q)+P_{8}(q,Q)\Sigma_{4}(q,Q)
\end{equation}
for the 6-loop case and 
\begin{equation}
P_{14}(q,Q)+P_{12}(q,Q)\Sigma_{2}(q,Q)+P_{10}(q,Q)\Sigma_{4}(q,Q)
+P_{8}(q,Q)\Sigma_{6}(q,Q)
\end{equation}
for 7-loop.

\subsection{Correction to the asymptotical BA}

The back reaction of the polarized vacuum is a change in the asymptotic
BA equation: 
\begin{equation}
pL+i\log S(p,-p)-2\pi=\Phi
\label{BAmod}
\end{equation}
where
\begin{equation}
\Phi=-\sum_{Q=1}^{\infty}\int\frac{dq}{2\pi}\mbox{sTr}((\partial_{q}S_{Q1}^{Q1}(q,p))
S_{Q1}^{Q1}(q,-p))e^{-\tilde{\epsilon}_{Q}(q)L}
\end{equation}
We can use eq. (\ref{BAmod}) to express $\delta p$ as 
\begin{equation}
\delta p=\frac{\Phi}{L+i\partial_{p}\log S(p,-p)}
\end{equation}
and with this correction the change in the energy is 
\begin{equation}
\Delta_{p}E=2\partial_{p}E(p_{ABA})\delta p
\end{equation}
As the derivative of the dispersion relaton is of order $g^{2}$ the
momentum shift shows up one order higher in the energy formula. Thus
it is sufficient to expand $\Phi$ up to the order $g^{12}$. 

We evaluated the differentiation in the integrand as follows:
\begin{eqnarray}
& & \mbox{sTr}((\partial_{q}S_{Q1}^{Q1}(q,p))S_{Q1}^{Q1}(q,-p))  = \\ 
 & & (\partial_{q} S_{\scriptstyle{scal}}(q,p))S_{scal}(q,-p)\mbox{sTr}(S_{mat}^{su(2)}(q,p)
S_{mat}^{su(2)}(q,-p))^{2}   \nonumber \\
& &   +2S_{scal}(q,p)S_{scal}(q,-p)\mbox{sTr}((\partial_{q}
S_{mat}^{su(2)}(q,p))S_{mat}^{su(2)}(q,-p)) \mbox{ sTr}
(S_{mat}^{su(2)}(q,p)S_{mat}^{su(2)}(q,-p))  \nonumber 
\end{eqnarray}
where 
\begin{eqnarray}
 \mbox{sTr}((\partial_{q}S_{mat}^{su(2)}(z,x^{\pm}))S_{mat}^{su(2)}(z,-x^{\mp}))= 
\phantom{aaaaaaaaaaaaaaaaaaaaaaaaaaaaaaaaaaaaaa}  & & \\
\phantom{aaaa} (Q+1)(\partial_{q}S1(z,x^{\pm}))S1(z,-x^{\mp}) 
+ (Q-1)(\partial_{q}S2(z,x^{\pm}))S2(z,-x^{\mp}) & &\nonumber \\
  -Q (\partial_{q}S3(z,x^{\pm}))S3(z,-x^{\mp})
-Q (\partial_{q}S4(z,x^{\pm}))S4(z,-x^{\mp}) &  &\nonumber 
\end{eqnarray}
In differentiating the scalar factor we need to differentiate the
$S_{0}$ part and the $\sigma$ part separately: 
\begin{equation}
\partial_{q}S_{scal}(q,p)=(\partial_{q}S_{0}(z,x^{\pm}))e^{i\sigma(z,x^{\pm})}
+S_{0}(z,x^{\pm})e^{i\sigma(z,x^{\pm})}(i\partial_{q}\sigma(z,x^{\pm}))
\end{equation}

\subsubsection*{Weak coupling expansion}

We expand the modification of the BA as 
\begin{equation}
\Phi=g^{8}\Phi_{8}+g^{10}\Phi_{10}+g^{12}\Phi_{12}+\dots
\end{equation}
which we write into the form 
\begin{equation}
\Phi=-\sum_{Q=1}^{\infty}\int\frac{dq}{2\pi}(dP(q,Q)\Sigma(q,Q)+P(q,Q)d\Sigma(q,Q))
\end{equation}
Besides notations used before we introduced 
\begin{eqnarray}
dP(q,Q) & = & (\partial_{q}S_{0}(q,p))S_{0}(q,-p)\mbox{sTr}(S_{mat}^{su(2)}(z,x^{\pm})
S_{mat}^{su(2)}(z,-x^{\mp}))^{2}+\\
 &  & 2S_{0}(q,p))S_{0}(q,-p)\mbox{sTr}((\partial_{q}S_{mat}^{su(2)}(z,x^{\pm}))
S_{mat}^{su(2)}(z,-x^{\mp}))\times \nonumber \\
& & \mbox{sTr}(S_{mat}^{su(2)}(z,x^{\pm}) 
S_{mat}^{su(2)}(z,-x^{\mp}))\nonumber 
\end{eqnarray}
and 
\begin{equation}
d\Sigma(q,Q)=i(\partial_{q}\sigma(z,x^{\pm}))\Sigma(q,Q)
\end{equation}
We need to expand each function to third nontrivial order
\begin{equation}
dP(q,Q)=dP(q,Q)_{8}g^{8}+dP(q,Q)_{10}g^{10}+dP(q,Q)_{12}g^{12}
\end{equation}
In the $\sigma$ case we need to keep only the first non-trivial term 
which survives after symmetrization:
\begin{equation}
\partial_{q}\sigma(z,x^{\pm})=\frac{g^{2}}{2}(\frac{1}{(x^{+})^{2}}-\frac{1}{(x^{-})^{2}})\partial_{q}(\Psi_{1}(1-igu^{+})-\Psi_{1}(1+igu^{+})+\Psi_{1}(1-igu^{-})-\Psi_{1}(1+igu^{-}))
\end{equation}

\section{Cross-checks for the integrand}

In the previous sections we have derived the integrand of the six and seven loop
contribution to the wrapping correction. This integrand consists of two
quite distinct parts -- the direct contribution to the energy which is
essentially just the transfer matrix $\eloop(q,Q)$ and the wrapping
contribution to the BY quantization condition $\emod(q,Q)$. This term
cannot be expressed in a simple form directly in terms of the transfer
matrix and we had to evaluate the derivatives of S-matrix
elements in order to calculate it (note that in contrast to the
relativistic case we cannot exchange the derivative w.r.t. the first
argument with minus the derivative w.r.t. the second argument).
We also symmetrized both expressions w.r.t. $q$.

The obtained expression has thus the form
\eq
\label{e.structure}
\Dl E_{F-term}= -\f{1}{2\pi} \sum_{Q=1}^\infty \int_{-\infty}^\infty dq\, (
\eloop(q,Q)-\emod(q,Q))
\eqx
The above expression is, strictly speaking, just the $F$-term L{\"u}scher
correction. $\mu$-term corrections could in principle arise from
the residues of the integrand at the dynamical poles ($s$- or $t$-channel
physical poles). However at weak coupling one can argue (see \cite{Bajnok:2008bm}) 
that $\mu$-terms should not appear and hence the residues at the dynamical poles
should cancel out after summation over $Q$. This indeed happened in our
earlier computations both at four and at five loops.
This constraint is a nontrivial cross check of the computation of both parts
of the integrand, since the sum of the residues turns out to vanish
only after we have included \emph{both} the direct energy contribution $\eloop(q,Q)$
and the BY modification contribution $\emod(q,Q)$.

Explicitly we find numerically that
\eq
\sum_{Q=1}^\infty \sum_{q_* \in \{q_5,q_6,q_7,q_8\}} {\rm res}_{q=q_*} \eloop(q,Q) =
\sum_{Q=1}^\infty \sum_{q_* \in \{q_5,q_6,q_7,q_8\}} {\rm res}_{q=q_*} \emod(q,Q)
\eqx
where the dynamical poles in the upper half plane are
\eq
q_{5,6} = \f{\mp 1}{\sqrt{3}}+i (Q-1) \qqqq
q_{7,8} = \f{\mp 1}{\sqrt{3}}+i (Q+1)
\eqx

Before we evaluate the full sum and integral numerically in the following section,
it is also important to derive an independent analytical cross-check for
our procedure of numerically fitting the coefficients of $\zeta$-functions
for the final expression.

Experience with earlier four- and five-loop computations show that one can
very easily derive analytically the coefficient of the single $\zeta$ function
of maximal transcendentality. It comes just from evaluating the integral
by residue on the `kinematical pole' $q=i Q$ ignoring any other poles coming
from the dressing phase. Moreover, this contribution arises only from the
rational part of the direct energy contribution $\eloop(q,Q)$.

Explicitly, we compute $-i\, {\rm res}_{q=iQ} P_{12}(q,Q)$, perform a partial
fraction expansion and look at the term proportional to $1/Q^9$. In the present
case we obtain
\eq
-i\, {\rm res}_{q=iQ} P_{12}(q,Q)= -\f{489 888}{Q^9}+\f{205 632}{Q^7} +\ldots
\eqx
We expect that the coefficient of the maximally transcendental $\zeta(9)$ is
$-489888$. We will not assume it in the subsequent computation but will check
at the end whether the numerically obtained coefficient of $\zeta(9)$ will
agree with this predicition. Note, however, that the coefficients of $\zeta$'s 
with lower transcendentality (like $\zeta(7)$ here) cannot be obtained in this way. 

Analogous computations for the seven loop integrand give
\eq
-i\, {\rm res}_{q=iQ} P_{14}(q,Q)= \f{7318080}{Q^{11}}+\ldots
\eqx
thus predicting that the coefficient of $\zeta(11)$ in the final answer
will be $7318080$. Again, we will not use this information in the numerical
fits, but will check whether this coefficient arises independently from the 
numerics. For the seven loop integrand we have also checked numerically
the cancellation of dynamical poles.

\section{The evaluation of the multiparticle L{\"u}scher correction}

We will now proceed to evaluate the expression (\ref{e.structure}). 
We will first give details for the six loop computation and then briefly
comment on the modifications at seven loops.

It is easy
to see that a natural variable for the integrand is $s=q/Q$ and one can see that
the integrand approaches at large $Q$ a single scaling function of $s$ (times
an appropriate power of $Q$). Hence we introduce
\eq
f_{6-loop}(s,Q)=Q (\eloop(s Q,Q)-\emod(s Q,Q))
\eqx
and evaluate
\eq
\label{e.sums}
-\sum_{Q=1}^\infty \int_0 ^\infty \f{ds}{\pi} f_{6-loop}(s,Q)
\eqx
We first numerically evaluate using Mathematica the integrals from $Q=1$ up 
to $Q_{max}=2000$
using 80 digit precision (we set {\tt PrecisionGoal}\footnote{With hindsight, 
setting {\tt AccuracyGoal} would be a more efficient option.} to 80 and {\tt WorkingPrecision}
to 100). By itself, this is a far too small $Q_{max}$ to estimate
the infinite sum with sufficient precision
in order to reliably fit the coefficients of the $\zeta$'s.
We therefore supplement this numerical part of the computation by computing
analytically the asymptotics of the integral around $Q=\infty$ to several orders
and performing the sums on these asymptotics from $Q=Q_{max}+1=2001$ to $Q=\infty$.

Explictly we find that the six-loop integral has the following asymptotic
expansion at large $Q$:
\eq
f_{6-loop}(s,Q)\sim \sum_{n=0} \f{1}{Q^{9+2n}} \left( a_n(s) + b_n(s) \log \f{1}{Q}+
c_n(s) \log^2 \f{1}{Q} \right)
\eqx
We evaluated the first six orders (up to $1/Q^{19}$) and integrated
the coefficient functions $a_n(s)$, $b_n(s)$, $c_n(s)$ (two of these integrals 
could be done analytically by Mathematica but 80-digit numerical evaluation
was faster and just as reliable). In this way we obtained the large $Q$ expansion
of the final summands
\eq
\label{e.summands}
\sum_{n=0}^5 \f{1}{Q^{9+2n}} \left( A_n + B_n \log \f{1}{Q}+
C_n \log^2 \f{1}{Q} \right)
\eqx
As a cross check, we verified that evaluating the above expression at our
cut-off point $Q=2000$ agrees with the numerical integral for the same $Q$ up
to $10^{-61}$.
Now we sum the asymptotic expression (\ref{e.summands}) from $Q=2001$ to 
$\infty$ and add it to the numerical sum.

We use EZ-Face \cite{EZFace} to express the resulting number as a linear combination
with integer coefficients. Using 44 digits precision we obtained
\eqn
\label{e.finalwrapsix}
\Dl E_{6-loop} &=& 261468 - 207360\, \zeta(3) + 156384\, \zeta(5) + 105840\, \zeta(7) - 
 489888\, \zeta(9) \nonumber \\
&& + 155520\, \zeta(3) \zeta(5) - 20736\, \zeta^2(3)
\eqnx 
We see that the the coefficient of $\zeta(9)$ obtained from the above semi-numerical
procedure exactly agrees with the analytical prediction of previous section.

In fact our numerical expression for $\Dl E$ obtained using $Q_{max}=2000$ and 
subsequent six orders of asymptotic expansion agrees with the above
analytical expression up to $10^{-59}$ (this is more than 60 decimal digits
agreement as $\Dl E=-42127.1014157...$). In order to make a further test, we
evaluated an additional seventh order of the asymptotic expansion. Incorporating
its contribution to the numerical expression for $\Dl E$ increased
its agreement with the analytical expression (\ref{e.finalwrapsix}) up to 
$10^{-66}$. Finally, let us note that if we were to keep only the numerical sum up to
$Q_{max} = 2000$ without the asymptotic expansions, the precision
would be much too small to fit the $\zeta$ functions. Indeed the deviation of this
sum from the exact answer starts already at order $10^{-20}$.

The evaluation of the seven loop wrapping correction proceeded along the same 
lines. We again evaluated the integrals numerically from $Q=1$ to $Q=2000$
using this time slightly different settings (we set now {\tt AccuracyGoal} to 70 
and {\tt WorkingPrecision} to 90). We could not increase these settings due to
some quirks in  Mathematica 8. Indeed, at the end of the computation we found that
Mathematica 8 did not evaluate the integrals for 24 values of $Q$, so we had
to evaluate these remaining integrals using Mathematica 7. 

Subsequently, we evaluated 9 orders of asymptotic expansion, the reason being 
that for the seven loop case we needed to fit 11 coefficients in contrast to 7 
at the six loop level, and hence we needed maximal numerical accuracy.
We thus evaluated
\eq
\sum_{n=0}^8 \f{1}{Q^{9+2n}} \left( A_n + B_n \log \f{1}{Q}+
C_n \log^2 \f{1}{Q} + D_n \log^3 \f{1}{Q} \right)
\eqx
in a similar manner as outlined above. The above expression evaluated
for $Q=1999$ agreed with the numerical integral for that value of $Q$
up to $10^{-75}$.
We then summed the above asymptotic expression
from $Q=2001$ to $Q=\infty$ and added the result to the sum of the numerical
integrals computed earlier.

In order to fit the coefficients of the $\zeta$'s, it is advantageous
to reduce the magnitude of the coefficients by factoring out
an integer divisor. Experience with lower loop levels suggests to factor
out 36. Now 65-digit precision in EZ-Face \cite{EZFace}, allowed us to obtain
the analytical answer:
\eqn
\label{e.finalwrapseven}
\Dl E_{7-loop} = 36 \cdot \biggl( &-& 139230+ 125904\, \zeta(3) - 22472\, \zeta(5) 
- 47964\, \zeta(7) - 32616\, \zeta(9) +\nonumber \\
& + & 203280\, \zeta(11) - 11808\, \zeta^2(3) + 11424\, \zeta(3) \zeta(5)
 - 53760 \, \zeta(3) \zeta(7) + \nonumber \\
& + &3456\, \zeta^3(3) -27600\, \zeta^2(5) \biggr)
\eqnx 
The deviation of the above expression from our semi-numerical answer appeared
only at order $10^{-73}$, therefore right at the edge of our numerical precision.
However, a key test of the above result is the coefficient of $\zeta(11)$
which was fitted to be $36 \cdot 203280 = 7318080$ and exactly coincides
with our analytical prediction from the previous section.

\section{The final result and conclusions}

In order to compute the final anomalous dimension of the Konishi operator,
it remains to add the six and seven loop wrapping corrections (\ref{e.finalwrapsix})
and (\ref{e.finalwrapseven}) to the relevant terms in the Asymptotic Bethe Ansatz
answer (\ref{e.ABA}). For completeness, we will present here the full result
up to seven loops, incorporating also the four loop wrapping correction
obtained in~\cite{Bajnok:2008bm}
\eq
\Dl E_{4-loop}=324 + 864\, \zeta(3) - 1440\, \zeta(5)
\eqx
and the five loop one from \cite{Bajnok:2009vm}
\eq
\Dl E_{5-loop}=-11340 + 2592\, \zeta(3) - 5184\, \zeta^2(3) - 11520\, \zeta(5) 
+ 30240\, \zeta(7)
\eqx
The final answer for the Konishi dimension is
\eqn
\Delta&=&4+12g^2-48g^4+336 g^6+ 96 (-26 + 6\, \zeta(3) - 15\, \zeta(5)) g^8 
 \\ \nonumber
&& -96 (-158 - 72\, \zeta(3) + 54\, \zeta(3)^2 + 90\, \zeta(5) - 315\, \zeta(7)) g^{10} 
\nonumber \\
&& -48 (160 + 432\, \zeta^2(3) - 2340\, \zeta(5) - 
    72\, \zeta(3) (-76 + 45\,  \zeta(5)) - 1575\,  \zeta(7) + 
    10206\,  \zeta(9))  g^{12} \nonumber\\
&& +48 (-44480 - 8784\, \zeta^2(3) + 2592\, \zeta^3(3) - 4776\, \zeta(5) - 
   20700\, \zeta^2(5) \nonumber \\  && + 24\,  \zeta(3) (4540 + 357\, \zeta(5) - 1680\,  \zeta(7)) 
 -   26145\,  \zeta(7) - 17406\, \zeta(9) + 152460\, \zeta(11)) g^{14}\nonumber
\eqnx
Our six loop result agrees with the computation of \cite{LSV} obtained from
FiNLIE thus being a very nontrivial cross check of the proposed finite set
of nonlinear equations describing the spectrum of $\nn=4$ SYM. It would be very 
interesting to compare this result with a direct gauge theoretical computations
along the lines of \cite{Eden:2012fe}. It would be also very interesting to
analytically verify our seven loop result, as in this case the identification
was performed at the edge of computational precision (however with a nontrivial
consistency check of the analytically known coefficient of $\zeta(11)$).
Finally let us note that extending the result for Konishi to eight loop order
would be conceptually quite important, as at that order double wrapping
effects will appear. These phenomena were analyzed for the  
nonprotected vacuum state in the $\gamma$-deformed theories in \cite{Ahn:2011xq}
but they are beyond the reach
of the L{\"u}scher correction formulas for excited states known up to now.

\bigskip

\noindent{\bf Acknowledgments:} We thank Gregory Korchemsky for suggesting 
this project to us. We are grateful to the authors of \cite{LSV}
for comunicating to us their result prior to publication, and to Dima Volin,
Kolya Gromov and Sergey Frolov for discussions leading to the elimination
of a discrepancy between our results. ZB thanks for the hospitality during 
the  CQUeST-IEU Focus program on
Finite-size Technology in Low Dimensional Quantum System (VI). ZB was 
supported by an MTA Lend\"ulet grant and by OTKA K81461.


\begin{thebibliography}{99}


\bibitem{BS}
  N.~Beisert and M.~Staudacher,
  ``Long-range PSU(2,2$|$4) Bethe ansaetze for gauge theory and strings,''
  Nucl.\ Phys.\ B {\bf 727}, 1 (2005)
  [hep-th/0504190].

\bibitem{AJK}
  J.~Ambjorn, R.~A.~Janik and C.~Kristjansen,
  ``Wrapping interactions and a new source of corrections to the spin-chain  /
  string duality,''
  Nucl.\ Phys.\  B {\bf 736}, 288 (2006)
  [arXiv:hep-th/0510171].

\bibitem{FSSZ}
  F.~Fiamberti, A.~Santambrogio, C.~Sieg and D.~Zanon,
  ``Wrapping at four loops in N=4 SYM,''
  Phys.\ Lett.\ B {\bf 666} (2008) 100
  [arXiv:0712.3522 [hep-th]].


\bibitem{Velizhanin08}
  V.~N.~Velizhanin,
  ``The four-loop anomalous dimension of the Konishi operator in N=4 supersymmetric Yang-Mills theory,''
  JETP Lett.\  {\bf 89} (2009) 6
  [arXiv:0808.3832 [hep-th]].


\bibitem{Bajnok:2008bm}
  Z.~Bajnok and R.~A.~Janik,
  ``Four-loop perturbative Konishi from strings and finite size effects for
  multiparticle states,''
  Nucl.\ Phys.\  B {\bf 807} (2009) 625
  [arXiv:0807.0399 [hep-th]].

\bibitem{Bajnok:2009vm}
  Z.~Bajnok, A.~Hegedus, R.~A.~Janik and T.~Lukowski,
  ``Five loop Konishi from AdS/CFT,''
  Nucl.\ Phys.\ B {\bf 827} (2010) 426
  [arXiv:0906.4062 [hep-th]].

\bibitem{Bajnok:2010ud} 
  Z.~Bajnok and O.~el Deeb,
  ``6-loop anomalous dimension of a single impurity operator from AdS/CFT and multiple zeta values,''
  JHEP {\bf 1101}, 054 (2011)
  [arXiv:1010.5606 [hep-th]].


\bibitem{TBA1}
  N.~Gromov, V.~Kazakov, A.~Kozak, P.~Vieira,
  ``Exact Spectrum of Anomalous Dimensions of Planar $\mathcal{N} = 4$ Supersymmetric Yang-Mills Theory: TBA and excited states,''
  Lett.\ Math.\ Phys.\  {\bf 91} (2010) 265,
  [arxiv:0902.4458 [hep-th]]


\bibitem{TBA2}
  D.~Bombardelli, D.~Fioravanti, R.~Tateo,
  ``Thermodynamic Bethe Ansatz for planar AdS/CFT: A Proposal,''
  J.\ Phys.\ A: Math.\ Theor.\ {\bf 42} (2009) 375401,
  [arxiv:0902.3930 [hep-th]]


\bibitem{TBA3}
  G.~Arutyunov, S.~Frolov,
  ``Thermodynamic Bethe Ansatz for the $AdS_5\times S^5$ Mirror Model,''
  JHEP {\bf 0905} (2009) 068,
  [arxiv:0903.0141 [hep-th]]

\bibitem{Arutyunov:2010gb}
  G.~Arutyunov, S.~Frolov and R.~Suzuki,
  ``Five-loop Konishi from the Mirror TBA,''
  JHEP {\bf 1004} (2010) 069
  [arXiv:1002.1711 [hep-th]].

\bibitem{Balog:2010xa}
  J.~Balog and A.~Hegedus,
  ``5-loop Konishi from linearized TBA and the XXX magnet,''
  JHEP {\bf 1006} (2010) 080
  [arXiv:1002.4142 [hep-th]].

\bibitem{transcend}
  A.~V.~Kotikov and L.~N.~Lipatov,
  ``DGLAP and BFKL equations in the N=4 supersymmetric gauge theory,''
  Nucl.\ Phys.\ B {\bf 661} (2003) 19
   [Erratum-ibid.\ B {\bf 685} (2004) 405]
  [hep-ph/0208220].

\bibitem{KLRSV}
  A.~V.~Kotikov, L.~N.~Lipatov, A.~Rej, M.~Staudacher and V.~N.~Velizhanin,
  ``Dressing and wrapping,''
  J.\ Stat.\ Mech.\  {\bf 0710} (2007) P10003
  [arXiv:0704.3586 [hep-th]].


\bibitem{fourlooptwisttwo}
  Z.~Bajnok, R.~A.~Janik and T.~Lukowski,
  ``Four loop twist two, BFKL, wrapping and strings,''
  Nucl.\ Phys.\ B {\bf 816} (2009) 376
  [arXiv:0811.4448 [hep-th]].

\bibitem{fivelooptwisttwo}
  T.~Lukowski, A.~Rej and V.~N.~Velizhanin,
  ``Five-Loop Anomalous Dimension of Twist-Two Operators,''
  Nucl.\ Phys.\ B {\bf 831} (2010) 105
  [arXiv:0912.1624 [hep-th]].

\bibitem{Gromov:2009zb}
  N.~Gromov, V.~Kazakov and P.~Vieira,
  ``Exact Spectrum of Planar ${\cal N}=4$ Supersymmetric Yang-Mills Theory: Konishi Dimension at Any Coupling,''
  Phys.\ Rev.\ Lett.\  {\bf 104} (2010) 211601
  [arXiv:0906.4240 [hep-th]].



\bibitem{Frolov:2010wt}
  S.~Frolov,
  ``Konishi operator at intermediate coupling,''
  J.\ Phys.\ A A {\bf 44} (2011) 065401
  [arXiv:1006.5032 [hep-th]].

\bibitem{Gromov:2011de}
  N.~Gromov, D.~Serban, I.~Shenderovich and D.~Volin,
  ``Quantum folded string and integrability: From finite size effects to Konishi dimension,''
  JHEP {\bf 1108} (2011) 046
  [arXiv:1102.1040 [hep-th]].

\bibitem{Roiban:2011fe}
  R.~Roiban and A.~A.~Tseytlin,
  ``Semiclassical string computation of strong-coupling corrections to dimensions of operators in Konishi multiplet,''
  Nucl.\ Phys.\ B {\bf 848} (2011) 251
  [arXiv:1102.1209 [hep-th]].

\bibitem{Vallilo:2011fj}
  B.~C.~Vallilo and L.~Mazzucato,
  ``The Konishi multiplet at strong coupling,''
  JHEP {\bf 1112} (2011) 029
  [arXiv:1102.1219 [hep-th]].

\bibitem{Eden:2012fe}
  B.~Eden, P.~Heslop, G.~P.~Korchemsky, V.~A.~Smirnov and E.~Sokatchev,
  ``Five-loop Konishi in N=4 SYM,''
  Nucl.\ Phys.\ B {\bf 862} (2012) 123
  [arXiv:1202.5733 [hep-th]].

\bibitem{NLIE1}
  N.~Gromov, V.~Kazakov, S.~Leurent, D.~Volin,
  ``Solving the AdS/CFT Y-system,''
  [arXiv:1110.0562 [hep-th]].

\bibitem{NLIE2}
  J.~Balog, A.~Hegedus,
  ``Hybrid-NLIE for the AdS/CFT spectral problem,''
  [arXiv:1202.3244 [hep-th]].
  
\bibitem{LSV} 
  S.~Leurent, D.~Serban and D.~Volin,
  ``Six-loop Konishi anomalous dimension from the Y-system,''
  arXiv:1209.0749 [hep-th].

\bibitem{Arutyunov:2008zt}
  G.~Arutyunov and S.~Frolov,
  ``The S-matrix of String Bound States,''
  Nucl.\ Phys.\ B {\bf 804} (2008) 90
  [arXiv:0803.4323 [hep-th]].

\bibitem{Arutyunov:2009kf}
  G.~Arutyunov and S.~Frolov,
  ``The Dressing Factor and Crossing Equations,''
  J.\ Phys.\ A A {\bf 42} (2009) 425401
  [arXiv:0904.4575 [hep-th]].

\bibitem{EZFace} See {\tt http://oldweb.cecm.sfu.ca/projects/EZFace/} 

\bibitem{Ahn:2011xq}
  C.~Ahn, Z.~Bajnok, D.~Bombardelli and R.~I.~Nepomechie,
  JHEP {\bf 1112} (2011) 059
  [arXiv:1108.4914 [hep-th]].


\end{thebibliography}
\end{document}